\documentclass[12pt]{iopart}

\usepackage{graphicx}
\usepackage{iopams}
\expandafter\let\csname equation*\endcsname\relax
\expandafter\let\csname endequation*\endcsname\relax
\usepackage{amsmath}
\usepackage{color} 
\usepackage{url}

\def\cO{\mathcal{O}}
\def \Ordo {\cO}
\def\ordo{o}

\begin{document}
\title{Modelling bursty time series}
\author{Szabolcs Vajna$^{1}$, B\'{a}lint T\'{o}th$^{2,3}$, J\'{a}nos Kert\'{e}sz$^{1,4}$}
\address{$^{1}$ Institute of Physics, BME, Budapest, Budafoki u. 8., H-1111}
\address{$^{2}$ School of Mathematics, University of Bristol, BS8 1TW}
\address{$^{3}$ Institute of Mathematics, BME, Budapest, Egry J\'{o}zsef u. 1., H-1111}
\address{$^{4}$ Center for Network Science, CEU, Budapest, N\'{a}dor u. 9., H-1051}
\ead{szabolcs.vajna@gmail.com} 

\begin{abstract}
Many human-related activities show power-law decaying interevent time distribution with exponents usually varying between $1$ and $2$. We study a simple task-queuing model, which produces bursty time series due to the nontrivial dynamics of the task list. The model is characterised by a priority distribution as an input parameter, which describes the choice procedure from the list. 
We give exact results on the asymptotic behaviour of the model and we show that the interevent time distribution is power-law decaying for any kind of input distributions that remain normalisable in the infinite list limit, with exponents tunable between $1$ and $2$. 
The model satisfies a scaling law between the exponents of interevent time distribution ($\beta$) and autocorrelation function ($\alpha$): $\alpha + \beta = 2$. This law is general for renewal processes with power-law decaying interevent time distribution. We conclude that slowly decaying autocorrelation function indicates long-range dependency only if the scaling law is violated.
\end{abstract}

\maketitle{}  

\section{Introduction} \label{sec:intro}
Studying human activity patterns is of central interest due to the wide practical usage. Understanding the dynamics underlying the timing of various human activities -- such as starting a phone call, sending an e-mail or visiting a web-site -- are crucial to modelling the spreading of information or viruses \cite{Vazquez_spreadproc}. Modelling human dynamics is also important in resource allocation. It has been shown that for many human activities the interevent time distribution follows a power law with exponents usually varying between $1$ and $2$. 
Processes with power-law decaying interevent time distribution look very different from the Poisson process, which has been used to model the timing of human activities before \cite{greene97,reynolds03}. While time series from the latter look rather homogeneous, the former processes produce bursts of rapidly occurring events, which are separated by long inactive periods.

Some examples where power-law decaying interevent time distribution has been observed are email-communication (with exponent $\beta\approx 1$, \cite{Eckmann2004}), surface mail communication ($\beta\approx 1.5$, \cite{OliBA2005}), web-browsing ($\beta \approx 1.2$, \cite{webbrows}) and library loans ($\beta\approx 1$, \cite{libloan}). In some other cases a monotonous relation has been reported between the user activity and the interevent time exponent, for example in short-message communication ($\beta \in (1.5,2.1)$, \cite{sms}) or in watching movies ($\beta\in(1.5,2.7)$, \cite{moviewatch}).
In a recent paper there can be found a distribution of exponents of various channels of communications \cite{exp_dist}. 
These observations make it necessary to find a model in which the interevent time exponent is tunable. 

Similar bursty behaviour has been observed in other natural phenomena, for example in neuron-firing sequences \cite{neuron} or in the timings of earthquakes \cite{earthquake}. The interevent time distribution does not give us any information about the dependency among the consecutive actions. Correlation between events is usually characterised by the autocorrelation function of the timings of detected events. Bursty behaviour is often accompanied by power-law decaying autocorrelation function \cite{Karsai_univ}, which is usually thought to indicate long-range dependency, see e.g. \cite{AutoDep}. However, time series with independent power-law distributed interevent times show long-range correlations \cite{Brain,IntCorr}. 

This paper is organized as follows. We start with introducing a task-queuing model, which has an advantage compared to the Barab\'{a}si-model \cite{BA2005}, namely that the observable is not the waiting time of an action (from adding the task to the list until executing it) 
but the interevent time between similar activities. We determine the asymptotic decay of the interevent time distribution in a simple limit of the model. We give a simple proof of the scaling law between the exponents of the interevent time distribution and the autocorrelation function based on Tauberian theorems. 
Finally, we demonstrate that the scaling law can be violated if the interevent times are long-range dependent.

\section{The model} \label{sec:model}
\begin{figure}[htb!] \centering
\includegraphics[width=10.0cm]{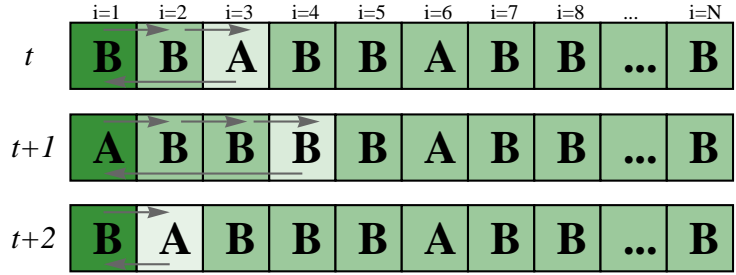}
\caption{Dynamics of the list. 
In every timestep a random position is chosen from the list and the task at the chosen position jumps to the first position initiating the corresponding type of activity. The other tasks are shifted to fill in the gap.
}
\label{fig: listdyn}
\end{figure}

We assume that people have a task list of size $N$ and they choose their forthcoming activity from
this list. The list is ordered in the sense that the probability $w_i$ of choosing the $i^{th}$ activity from the
list is 
decreasing as a function of position $i$. 
The task at the chosen position jumps to the front (triggering the corresponding activity) and pushes the tasks that preceded it to the right (figure \ref{fig: listdyn}).
This mechanism is responsible for producing the bursty behaviour because  
once a person starts to do an activity, that is going to have high priority for a while.  

The model is capable of covering many types of activities but now we only concentrate on one type. The tasks of this type are marked with $A$ in figure \ref{fig: listdyn} (e.g. watching movies). For the sake of simplicity we assume that the list contains only one single item of type $A$, all the others are activities of type $B$. It turns out that this restriction is important for the interevent time distribution but irrelevant for the autocorrelation function.

We show in our paper that this model produces power-law decaying interevent time distribution for \emph{a wide variety of the $w_i$ distributions}. First we analyse the model with power-law decaying priority distribution $w_i$, second we analyse the effect of exponentially decaying priority distribution, and finally we discuss the stretched exponential case. 
If the list is finite, the power-law regime is followed by an exponential cutoff. The cutoff is the consequence of reaching the end of the list, from which a geometrically distributed waiting time follows: $P_{wt}(t) \sim (1-w_N)^t$. A marginal result of section \ref{sec:autocorr} shows that the expectation value of the interevent time is equal to the length of the list independently on the priority distribution (as long as ergodicity is maintained).
For the sake of simplicity we determine the exponent of power-law decaying region in the case of an infinitely long list where the exponential cutoff does not appear.

\section{Interevent time distribution}
\begin{figure}[htb!] \centering
\includegraphics[width=9.0cm]{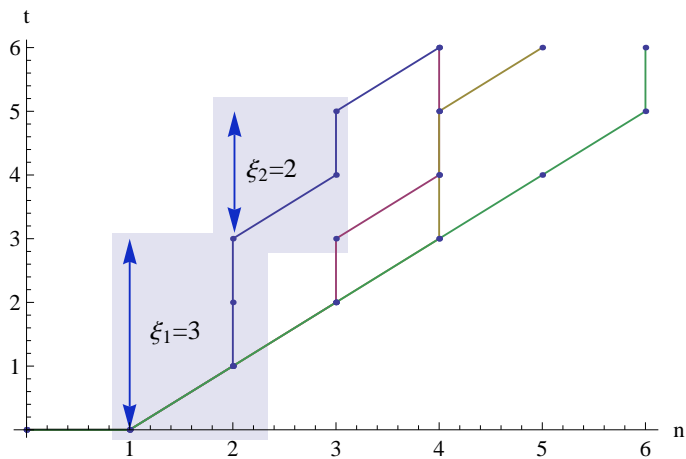}
\caption{Some examples for the path in the $(n,t)$ plane. Some typical sections in the path corresponding to the $\xi_k$ random variable are emphasized by a rectangle.}
\label{fig: q_nt_exp}
\end{figure}

The interevent time is equal to the recurrence time of the first element of the list. 
Let $q(n,t)$ $(n\geq1)$ denote the probability of finding the observed element at position $n$ after $t$ timesteps without any recurrences up to time $t$. We emphasise here that $q(n,t)$ is not a conditional probability and that $1-\sum_{n=1}^{\infty} q(n,t)$ gives the survival function of the interevent time distribution. The initial condition is set as $q(n,1) =(1-w_1) \delta_{n,2}$, that is, at the first step the observed element moves from the front of the list to the second position with probability $(1-w_1)$ (otherwise a recurrence would occur). The restriction not to recur is important because this makes large jumps to the front of the list \emph{forbidden} for the observed element. Time evolution is given by
\begin{eqnarray} \label{eq:q1} 
q(1,t) &\equiv 0 \\ \label{eq:qn}
q(n,t+1)&=(1-Q_{n-1})q(n,t)+Q_{n-1}q(n-1,t) \,,
\end{eqnarray}
where $Q_{n}=1-\sum_{k=1}^{n}w_k$ is the survival function of the priority distribution. Equation \eref{eq:qn} expresses that the observed element can be found at position $n$ either if it was there in the previous time step and we choose a position smaller than $n$ or it was at position $n-1$ and we choose an activity at position $k>n$.  
Our aim is to determine the asymptotic behaviour of $\sum_{n=1}^{\infty} q(n,t)$. The main trick we use in analysing the recursion above is to consider the $n=const$ levels first instead of the $t=const$ levels which intuitively one would do. 
At every step the time coordinate gets increased while the position of the observed element might remain unchanged or get increased by one as well (figure \ref{fig: q_nt_exp}).
The path of the element in the $(n,t)$ plane can be divided into sections that start with a step on the bias and are followed by some steps upwards (which can be zero as well). These sections can be characterised by their height-difference, which can take values from $1,2,\dots$. These height differences are independent and (optimistic) geometrically distributed with parameter depending on the position. Let $\xi_k$ be independent geometrically distributed random variables with parameter $Q_{k}$, i.e. $\mathbb{P}(\xi_i=\tau)=(1-Q_{i})^{\tau-1}Q_{i}$.
With $n$ fixed $q(n,t)$ is the probability that we find the element at the $n^{th}$ position at time $t$ (without any recurrences). This corresponds to paths with $n-1$ steps to the right (started from position $1$ at time $0$) and $q(n,t)$ is the distribution mass function of the sum of the random heights.
\begin{equation} \label{eq:qnt_xi}
q(n,t)=\mathbb{P}\left(\sum_{i=1}^{n-1}\xi_i = t\right)
\end{equation} 
We could go so far without specifying the priority distribution, now we have to specify $Q_n$ to continue.   

\subsection{Power-law decaying priorities}
\begin{figure}[htb!] \centering
\includegraphics[width=6.0cm]{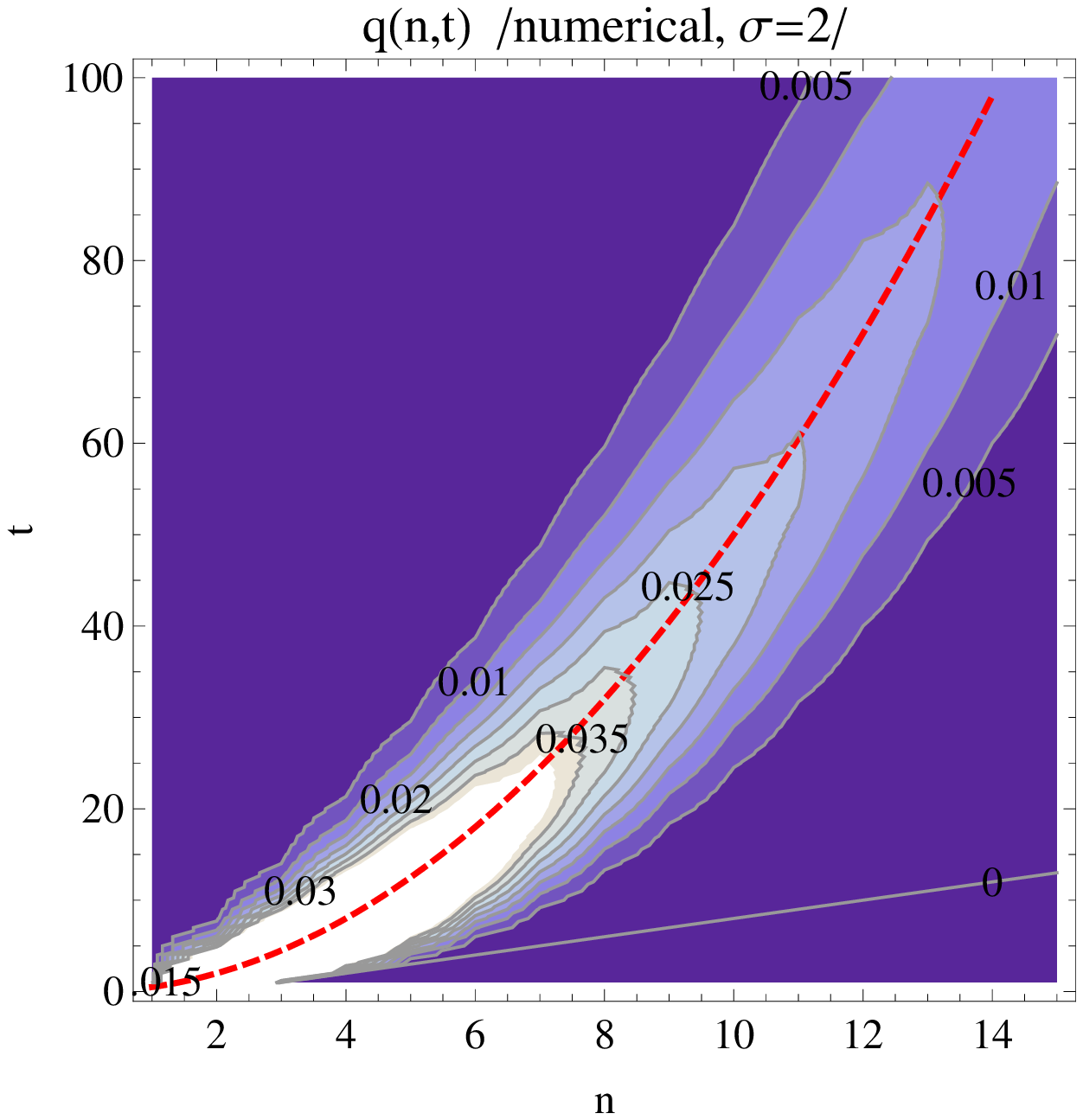}
\includegraphics[width=6.0cm]{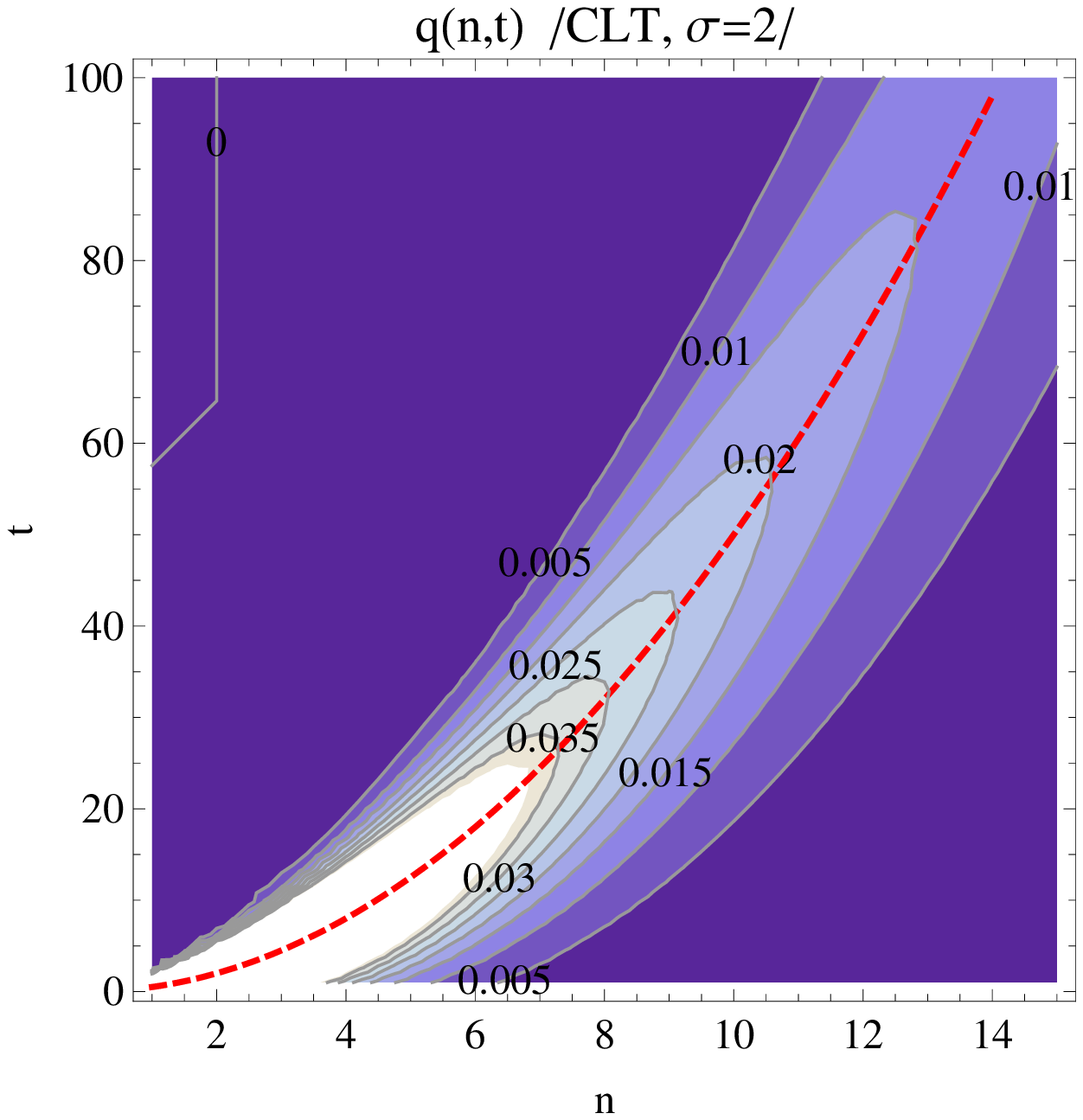}
\caption{Contour plots of $q(n,t)$. The image on the left is a numerical result calculated directly form the recursion equations (\ref{eq:q1}-\ref{eq:qn}), the right plot shows the CLT approximation \eref{eq: q_nt_b}. The dashed curve is $t=\frac{n^{\sigma}}{\sigma}$ ($c=1$).}
\label{fig: q_nt_cont}
\end{figure}

Here we analyse the model in which the survival function of the priority distribution is power-law decaying, i.e.  $Q_{n}= c n^{-\sigma+1} + \Ordo(n^{-\sigma})$. 
$\Ordo(n^{\omega})$ means that that term is asymptotically at most of the order of $n^{\omega}$. In this case $w_i$ is also power-law decaying with exponent $\sigma$. 
Though the random variables $\xi_k$ are not identically distributed -- by checking the Lyapunov condition -- one can show 
that the central limit theorem holds for this situation (see \ref{sec:appCLTpower}). 
In this approximation $q(n,t)$ is Gaussian in variable $t$ with mean $\sum_{i=1}^{n-1}\mathbb{E}\left(\xi_i\right)$ and variance $\sum_{i=1}^{n-1}\mathbb{D}^2\left(\xi_i\right)$, where $\mathbb{E}\left(\xi_i\right)=Q_i^{-1}$ and $\mathbb{D}^2\left(\xi_i\right)=Q_i^{-2}(1-Q_i)$. Using integral approximation to evaluate the sums and keeping only the highest order terms in $n$ yields:
\begin{equation} \label{eq: q_nt_b}
q(n,t) \stackrel{CLT}{\approx} \frac{c\sqrt{{2\sigma-1}}}{\sqrt{2 \pi} n^{\sigma-1/2}}  
\exp \left\{ -\frac{c^2(2\sigma-1)}{2} \frac{\left(t-\frac{n^{\sigma}}{c\sigma}\right)^2}{n^{2\sigma-1}} \right\} \,.
\end{equation}
This formula shows that the probability of finding an element at position $n$ at time $t$ is centered on the $t=\frac{n^{\sigma}}{c\sigma}$ curve which is in agreement with the numerical results (figure \ref{fig: q_nt_cont}).

The sum of $q(n,t)$ in variable $n$ is the probability of the observed element has not recurred up to time $t$. 
We approximate this sum by integral and we apply the following substitution (with new variable $r$):
\begin{equation}
n=(c\sigma t)^{1/\sigma} - r (c\sigma t) ^{1/(2\sigma)} \,.
\end{equation}
For $\gamma>0$ (here $\gamma$ may refer to: $\sigma-1/2, \sigma, 2\sigma-1$):
\begin{equation} \label{eq: ngamma}
n^\gamma = (c\sigma t)^{\frac{\gamma}{\sigma}} - \gamma (c\sigma t)^{\frac{2\gamma-1}{2\sigma}} r + \ordo(t^{\frac{2\gamma-1}{2\sigma}})\, ,
\end{equation}
where $\ordo(n^{\omega})$ means that that term is asymptotically of strictly smaller
order than $n^{\omega}$. From equations (\ref{eq: q_nt_b}-\ref{eq: ngamma}) 
it follows that
\begin{equation}
\int_0^\infty  q(n,t) dn = c^{\frac{1}{\sigma}}(\sigma t)^{\frac{1}{\sigma}-1} + \ordo (t^{\frac{1}{\sigma}-1})\,. 
\end{equation}
Differentiating this equation gives the first main result of our paper, $P_{ie}(t) \sim t^{-\beta}$ with $\beta=2-\frac{1}{\sigma}$.

\subsection{Exponentially decaying priorities}
Now we study the model with $Q_n=c e^{-\lambda n}$ priority survival function. In contrast to the previous case the central limit theorem cannot be used, because the last term in $\sum_{i=1}^{n-1}\xi_i$ is comparable with the complete sum.

However, we can construct a limit theorem even for this situation. First of all, we approximate the geometrically distributed random variables $\xi_k \sim Geom(1-Q_k)$ by exponential ones: $\tilde{\xi}_k \sim Exp(Q_k)$. We apply the scaling property of the exponential distribution to express $\tilde{\xi}_k$ by i.i.d. exponential random variables: $\tilde{\xi}_k \sim e^{\lambda k} \eta_k$, where $\eta_k \sim Exp(c)$. In equation \eref{eq:qnt_xi} we need the distribution of sums of the first $n-1$ $\xi$s:
\begin{eqnarray} \label{eq:expscale}
&\sum_{i=1}^{n-1}\tilde{\xi}_i \sim e^{n \lambda} X_{n-1}\\
& X_{n-1}=\sum_{k=1}^{n} e^{-k\lambda}\eta_k \,.
\end{eqnarray} 
$X_n$ has a well defined limit as $n\rightarrow\infty$, which we denote by $X$. The probability density function of this (limit) random variable is denoted by $\psi(x)$ (and is given explicitly in \ref{sec:appLTexp}). The finite sum $X_n$ in equation \eref{eq:expscale} can be approximated by $X$, yielding $\mathbb{P}(\sum_{i=1}^{n-1}\tilde{\xi}_i < t)\approx \mathbb{P}(X < e^{-\lambda n}t) $ and
\begin{equation}
\int_{0}^{\infty} q(n,t) dn \approx \int_{0}^{\infty} e^{-\lambda n} \psi(t e^{-\lambda n}) dn = \frac{1}{\lambda t}\int_0^t\psi(y) dy =\frac{1}{\lambda t} + \ordo(t^{-1}) \,,
\end{equation}
since $\int_0^t\psi(y) dy$ tends to 1 as $t\rightarrow \infty$. By differentiating we get $\beta=2$ independently on parameter $\lambda$.

\subsection{Stretched exponentially decaying priorities}
The last investigated priority distribution is the stretched exponential, i.e. $Q_n=c e^{-\lambda n^{\nu}}$.
\subsubsection*{Faster than exponential decay ($\nu>1$)}
In this case $\sum_{i=1}^{n-1}{\xi_i}$ is totally dominated by the last term in the sum, hence we approximate the sum with this term.
\begin{equation}
q(n,t) \approx c e^{-\lambda n^{\nu}} e^{-t c e^{-\lambda n^{\nu}}}
\end{equation}
\begin{eqnarray}
\int_0^{\infty} q(n,t) dn &= \frac{1}{t\lambda^{1/\nu} \nu} \int_{0}^{t c} e^{-y} (\log(tc)-\log(y))^{-(1-1/\nu)} dy \\
&=\frac{1}{\lambda^{1/\nu} \nu} \frac{1}{t (\log(tc))^{1-1/\nu}}+ \ordo(\frac{1}{t (\log(tc))^{1-1/\nu}})
\end{eqnarray}
This means that the $\beta=2$ exponent holds for this case but a logarithmic correction appears. 

\subsubsection*{Slower than exponential decay ($\nu<1$)}
In this case the increment of $Q_n$ is just small enough for the central limit theorem to be still applicable (see appendix).
We approximate the mean and variance of $\sum_{i=1}^{n-1}\xi_i$ by integral and we use the property $\int_{0}^{a} e^{y} y^{1/\nu-1} \stackrel{a\rightarrow \infty}{\approx} e^{a} a^{1/\nu-1}$. With these we get
\begin{equation}
q(n,t) \approx \sqrt{\frac{c^2 \lambda \nu}{\pi}} e^{-\lambda n^{\nu}}n^{-\frac{1-\nu}{2}}
\exp\left\{-\frac{\left(t-e^{\lambda n^{\nu}}n^{1-\nu}/(c \lambda \nu)\right)^2}{2 \frac{1}{2 c^2\lambda \nu} e^{2\lambda n^{\nu}}n^{1-\nu}}\right\} \,.
\end{equation}
The $t\rightarrow \infty$ asymptotic behaviour of $\int q(n,t) dn$ can be determined by introducing a proper new variable,
\begin{equation} \label{eq:newvarsmexp}
y= n^{\frac{1-\nu}{2}}-c \lambda \nu t e^{-\lambda n^{\nu}}n^{\frac{\nu-1}{2}} \,.
\end{equation}
Keeping the leading order term only, one finds that the survival function of the interevent time distribution is
\begin{equation}
\int_{0}^{\infty} q(n,t) dn =\frac{1}{\lambda^{1/\nu}\nu} \frac{1}{t \log(c\lambda\nu t)^{1-1/\nu}}+\ordo(\frac{1}{t \log(c\lambda\nu t)^{1-1/\nu}}) \,.
\end{equation}
For more details of the calculations see \ref{sec:appCLTstrexp}. We have got $\beta=2$ with logarithmic correction again. While for $\nu > 1$ the correction makes the decay of the distribution faster than the pure power law, for $\nu < 1$ it is slightly slowed down.

\section{Autocorrelation function} \label{sec:autocorr}
\begin{figure}[htb!] \centering
\includegraphics[width=10.0cm]{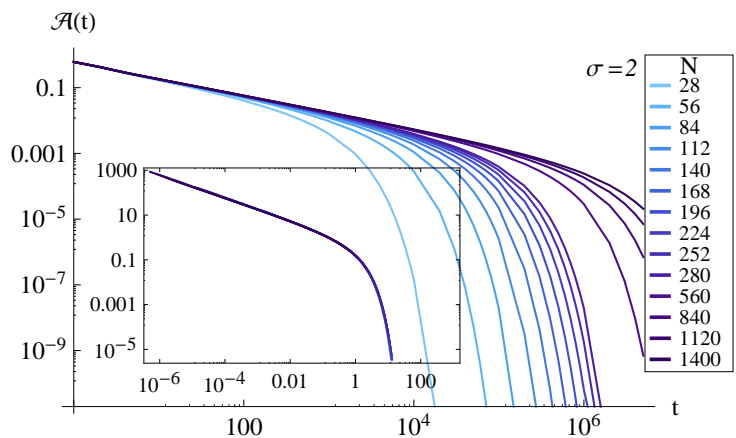}
\caption{Autocorrelation function of the model with various list sizes (with power-law decaying priority distribution). Inset: the curves corresponding to different $N$s collapse into a single curve if $N^{\delta}A(t)$ is plotted as a function of $t/N^{\gamma}$.}
\label{fig: auto_skala}
\end{figure}

Another characteristic property of the time series is autocorrelation function. Let $X(t)$ denote the indicator variable of the observed activity: $X(t)=1$ if the observed activity is at the first position of the list at time $t$ (i.e. an event happened) and $X(t)=0$ otherwise. We define the autocorrelation function as usual,
\begin{equation} \label{eq:autocorrdef}
\mathcal{A}(t)=\frac{\left\langle \mathbb{E}\left[X(s)X(t+s)\right]\right\rangle_s-\left\langle \mathbb{E}\left[X(s)\right] \right\rangle_s^2}{\left\langle \mathbb{E}\left[X(s)\right] \right\rangle_s-\left\langle \mathbb{E}\left[X(s)\right] \right\rangle_s^2}.
\end{equation} 
The model defines a Markov-chain and the state space is the space of permutations of the list. 
The transition matrix for a list of length $N$ is given by
\begin{equation} \label{eq:mastermatrix}
M=
\begin{pmatrix} 
1-Q_1		&Q_1			&0 		&\cdots &0\\
Q_1-Q_2 &1-Q_1 		& Q_2 & 0 		&\vdots\\
Q_2-Q_3 &0 				&1-Q_2&Q_3 		& 0\\
\vdots 	&\vdots 	&0 		&\ddots &Q_{N-1}\\
Q_{N-1}-Q_{N} &0&\cdots &0 &1-Q_{N-1}
\end{pmatrix} \, ,
\end{equation} 
where $Q_N=0$. This matrix is doubly stochastic, hence $\left\langle \mathbb{E}\left[X(t)\right]\right\rangle_t=\frac{1}{N}$. According to the recurrence formula of Marc Kac \cite{Kac} we conclude that \emph{the expectation value of the interevent time is equal to the size of the list independently on the priority distribution} (as long as ergodicity is maintained). 
The autocorrelation function can be written in simpler form making use of the knowledge of $\left\langle \mathbb{E}\left[X(t)\right]\right\rangle_t$,
\begin{equation} \label{eq:modellautocorrX} 
\mathcal{A}(t)=\frac{\mathbb{P}\left(X(t)=1 | X(0)=1\right)-\frac{1}{N}}{1-\frac{1}{N}} \,.
\end{equation} 
The probability of finding the activity at the first position can be calculated numerically by successive application of the Markov-chain transition matrix \eref{eq:mastermatrix}. 
For power-law decaying priority distributions numerical computations show that the autocorrelation function is power-law decaying with an exponential cutoff (figure \ref{fig: auto_skala}). Given $\sigma$, the autocorrelation functions for various list sizes can be rescaled to collapse into a single curve (figure \ref{fig: auto_skala}, inset).
This property can be written in a mathematical way:
\begin{equation} \label{eq:caaskala_trans}
\mathcal{A}(t,N,\sigma)=N^{-\delta} f_{\sigma} \left( \frac{t}{N^{\gamma}} \right)
=t^{-\frac{\delta}{\gamma}} \tilde{f}_{\sigma} \left( \frac{t}{N^{\gamma}} \right) \,.
\end{equation} 
The exponents used for rescaling the autocorrelation functions are listed in table \ref{tab: gammadelta}.
\begin{table}[htb] 
\center
\begin{tabular} {| l || *9{ c |}}
\hline
$\sigma$ &0.5 & 0.8 & 1 & 1.5 & 1.8 & 2 & 3 & 4 & 5\\ \hline \hline
$\gamma$ &1.07 &1.15 & 1.17 & 1.57 & 1.8 & 2 & 3 & 4 & 5\\ \hline
$\delta$ &1 & 1 & 1 & 1 & 1 & 1 & 1 & 1 & 1\\
\hline
\end{tabular}
\caption{Numerical results for data collapse in figure \ref{fig: auto_skala}: scaling parameters $\gamma$ and $\delta$ for various values of $\sigma$.}
\label{tab: gammadelta}
\end{table} 
indicating that in the $N\rightarrow \infty$ limit the autocorrelation function is power-law decaying with exponent $\alpha=\frac{1}{\sigma}$ for $\sigma \geq 2$. With the exact $\beta=2-\frac{1}{\sigma}$ result this can be combined into a scaling law: $\alpha+\beta=2$.

If the priority distribution decays faster than power-law, then $\beta=2$ and $\alpha=0$. This means that the autocorrelation function decays slower than any power. This is the case when the priority distribution decays like a (stretched) exponential function. 

\section{Proof of the scaling law}
The essential properties of the model for the scaling relation are that the interevent times are
independent and power-law decaying. Let $T$ denote the set of recurrence times and let $\tau$ be an interevent time. The autocorrelation function can be written in the following form:
\begin{equation}
\mathcal{A}(t)=\frac{\mathbb{P} \left( t \in T \right)-\frac{1}{\mathbb{E}\left[\tau\right]}}{1-\frac{1}{\mathbb{E}\left[\tau\right]}} \,,
\end{equation}
which simplifies to $\mathcal{A}(t)=\mathbb{P} \left( t \in T \right)$ if $\beta \leq 2$.
The Laplace transform of the autocorrelation function can be expressed by the Laplace transform of the interevent time distribution:
\begin{equation}
g(\lambda) \equiv \sum_{t=0}^{\infty}{\mathcal{A}(t) e^{-\lambda t}}=\left( 1-\mathbb{E} \left[ e^{- \lambda \tau} \right] \right)^{-1} ,
\end{equation}
where we used the property that the interevent times are independent and identically distributed. 
Tauberian and Abelian theorems connect the asymptotics of a function with the asymptotics of its Laplace transform \cite{Feller2}. Applying Abelian theorem to the right side of the last equation results in $g(\lambda)\sim \lambda^{1-\beta}$. Then applying the Tauberian theorem yields $\mathcal{A}(t) \sim t^{\beta-2}$ or
\begin{equation} \label{eq: scaleAB}
\alpha+\beta=2 \,.
\end{equation}
To be precise we had to use an extended version of the Abelian theorem, which can be derived from the original theorem using integration by parts. 
With similar train of thought the scaling law can be extended to the $2<\beta<3$ region where $\beta-\alpha=2$ holds. These results are in agreement with \cite{IntCorr,Autocorr_RenProc}.

\section{Models with non-independent interevent time distribution} \label{sec:nonindep}
Independence of the interevent times was important in the proof of the scaling law. The violation of the scaling relation indicates dependency among the interevent times but we emphasize that the opposite direction does not necessarily hold. In this section we investigate two models that are characterised by dependent interevent times. One of them satisfies the scaling relation \eref{eq: scaleAB} and the other does not.

\subsection{A model with weak dependence}
\begin{figure}[htb!] \centering
a) \includegraphics[width=6.0cm]{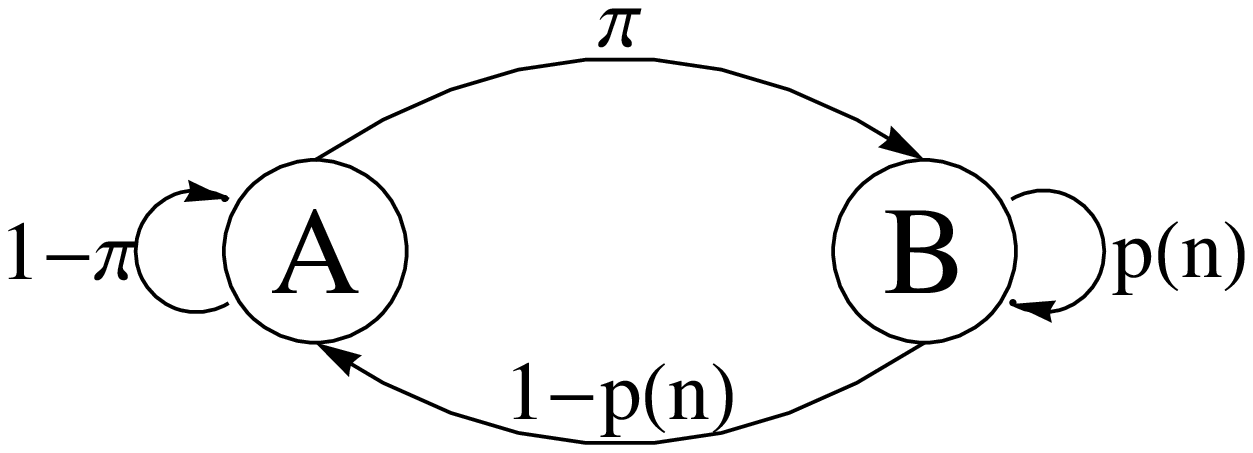}
b) \includegraphics[width=6.5cm]{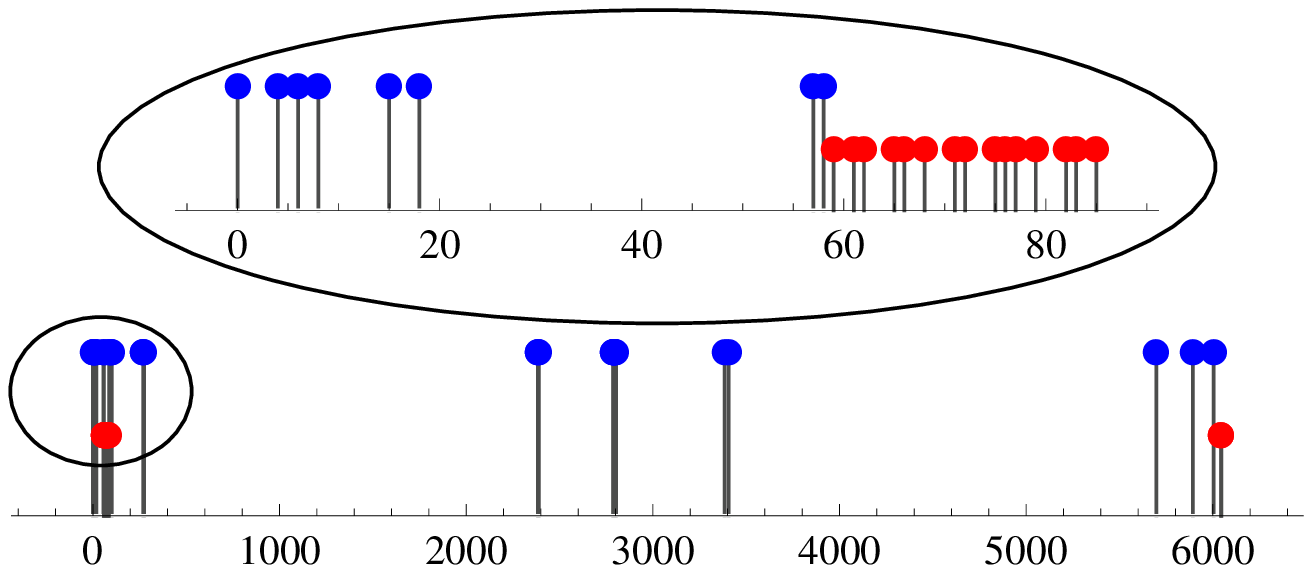}
\caption{a) Schematics of the two-state model. b) Time series generated by the two-state model. The higher sticks correspond to events performed in state $A$ and the lower sticks correspond to events performed in state $B$. Looking at the time series at a larger scale the ratio of $B$ periods shrinks. For the better visibility of the time series the model parameters were changed to $\beta_A=1.5$ and $\beta_B=3$.}
\label{fig: 2statemodel}
\end{figure}

Our first example is the two-state model of reference \cite{Karsai_univ}, because that model was introduced to capture deep correlations, nevertheless, it obeys the scaling law. We start with a brief introduction of the model. The system alternates between a normal state $A$, in which the events are separated by longer time intervals, and an excited state $B$, which performs events with higher frequency. After each event executed in state $A$ the system may switch to state $B$ with probability $\pi$ or remain in state $A$ with probability $1-\pi$. In contrast, the excited state has a memory. That is, if the system had executed more events in state $B$ since the last event in the normal state, the transition to state $A$ would be less probable. 
The probability of performing one more event in state $B$ is given by $p(n)=\left(n/(n+1)\right)^\nu$, where $n$ is the number of events in the current excited state. 
This memory function was introduced to  model the empirically observed power-law decay in the bursty event number distribution (for the definition see the reference). The dynamics is illustrated in figure \ref{fig: 2statemodel}a) for a better understanding. In both states the interevent times are generated from a reinforcement process resulting independent random interevent times $\tau_A$ and $\tau_B$ with power-law decaying distributions. We denote the corresponding exponents by $\beta_A$ and $\beta_B$. The parameters of the model were the following: $\pi=0.1$, $\nu=2.0$, $\beta_A=1.3$, $\beta_B=5.0$, and it was  found that the model satisfies the scaling law with exponents $\beta=1.3$ and $\alpha=0.7$.

The number of events performed in a single period of state $A$ or $B$ are i.i.d. random variables $N_A$ or $N_B$. $N_A$ is geometrically distributed with parameter $\pi$. The distribution of $N_B$ can be easily obtained from $p(n)$, and it is power-law decaying with exponent $1+\nu$. Making use of these new random variables the dynamics simplifies: the system executes $N_A$ events with interevent times distributed as $\tau_A$, then switches to the excited state and executes $N_B$ events with interevent times $\tau_B$, then switches back to the normal state, and so on.

It is clear that the decay of the interevent time distribution of the whole process is determined by the smallest of the exponents $\beta_A$ and $\beta_B$. Since both of $\mathbb{E}\left[\tau_B\right]$ and $\mathbb{E} \left[N_B\right]$ are finite, the periods of state $B$ have a characteristic length $\xi_t=\mathbb{E}\left[\tau_B\right]\mathbb{E}\left[N_B\right]$. Within the periods of state $A$ there is no characteristic time-scale, hence by looking on the time series at a larger scale the effective length of the $B$ periods will be shorter (figure \ref{fig: 2statemodel}b)). Therefore the existence of the $B$ periods is irrelevant for the asymptotic decay of the autocorrelation function giving rise to the scaling law.

\subsection{A model with long-range dependence} 
Another simple way to introduce dependency between the interevent times is to construct a Markov-chain, i.e. the next interevent time depends only on the actual interevent time.  As a counterexample to the scaling law 
we constructed a long-range dependent set of interevent times by Metropolis-algorithm. 
The base of the algorithm is constructing a Markov chain on the integers that has power-law decaying stationary distribution $P(x)\sim x^{-\beta}$. 
The algorithm uses a proposal density (transition rate) $Q(x',x_n)$ which generates a proposal sample from the current value of the interevent time. This sample is accepted for the next value with probability
\begin{equation}
\alpha(x',x_n)=\min\left\{\frac{Q(x_n,x')P(x')}{Q(x',x_n)P(x_n)},1\right\} \, ,
\end{equation}
otherwise the previous value is repeated.
To generate a power-law distributed sample with long-range dependency the mixing of the Markov chain should be slow, i.e. the gap in the spectrum of the Markov chain should vanish. 
In this order we allow only small differences between consecutive interevent times: 
$Q(x',x_n)=\frac{1}{2D}\mathbb{I}(x_n-D,x_n+D)$ or equivalently $x'$ as a random variable is discrete uniformly distributed: $x'\sim DU[x_n-D,x_n+D]$.
\begin{figure}[htb!] \centering
\includegraphics[width=10.0cm]{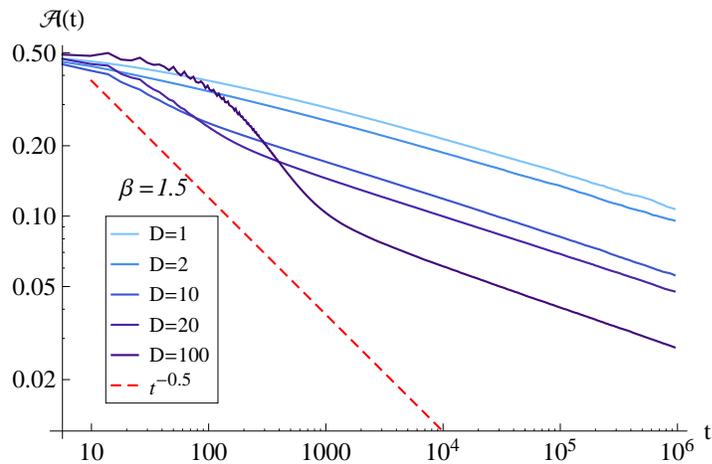}
\caption{Simulation results of the Metropolis algorithm. The exponent of the interevent time distribution is set to $\beta=1.5$ and parameter $D$ of the proposal density is varied. The dashed line shows the decay corresponding to the scaling law of independent processes.}
\label{fig: metro}
\end{figure} 
The simulation results (figure \ref{fig: metro}) show that the autocorrelation function decays slower than it would be assumed from the scaling law \eref{eq: scaleAB}. 
This example shows that the scaling relation \eref{eq: scaleAB} can be violated if there is long range dependency among the interevent times.

\section{Extensions of the model}

A trivial extension of the model could be putting more items of the observed activity into the list. Then this parameter would tune the frequency of doing the activity. In this case the interevent times become dependent and the simulation results on finite lists show that the interevent time distribution decays faster than power-law but decays slower than an exponential function. It is easy to prove that in spite of this the autocorrelation function \eref{eq:autocorrdef} is independent of the number of the observed activity and remains power-law decaying.  This is another example for models in which the scaling law does not hold.

When the list is finite, we have much more freedom in the choice of the priority distribution. 
However, the method based on expressing $q(n,t)$ as a probability density function of sum of $n-1$ independent geometrically distributed random variables is general, and can offer a good base for further calculations, even for finite lists. 

The model may be interesting not only for human dynamics but also for the mathematical theory of card shuffling \cite{Diaconis}.  
We can define the time reversed version of the model in which we choose a position from the same distribution as in the original model and we move the first element of the list to the random position. This model has similar properties to the original model, e.g. this model has the same interevent time distribution and autocorrelation function. If we think of the list as a \emph{deck of cards}, then the time reversed model is a generalisation of the \emph{top-in-at-random shuffle} 
method. The latter is a primitive model of shuffling cards: the top card is removed and inserted into the deck at a uniformly distributed random position \cite{Diaconis}.

\section{Conclusion}

In a proper model one should take into consideration the dependency among the interevent times besides the interevent time distribution. In human dynamics the latter is slowly decaying and as a consequence of this the autocorrelation function of the interevent times is not well-defined (i.e. the second moment does not exist). However, the autocorrelation function of the time series exists and it might be a good measure for long range dependency among the interevent times. 
Time series of messages sent by an individual in an online community are reported to be not more correlated than an independent process with the same interevent time distribution \cite{IntCorr}. Similarly, the exponents in the neuron-firing sequences approximately satisfy the scaling law ($\alpha=1.0$, $\beta=1.1$ \cite{Karsai_univ}). For these systems the model we studied might be applicable. However, this is not always the case. 
For example, the autocorrelation function of the e-mail sequence decays slower ($\alpha=0.75$ \cite{Karsai_univ}) than it should do estimated from the scaling law. This indicates long range dependency in the time series in addition to the power-law decaying interevent time distribution. 
In this case a dependent model should be applied, for example a model based on Metropolis algorithm that is similar to the one we studied as a counterexample to the scaling law. Effects of circadian patterns and inhomogeneity in an ensemble of individuals should also be considered \cite{KimCirc}.

In real networks interactions between individuals have to be taken into account 
to reproduce some social phenomena, e.g. temporal motifs \cite{TempMotif} observed in a mobile call dataset.  Interactions could be incorporated in this model by allowing the actual activity of an individual to modify the priority list of some of his/her neighbour. If this effect is rare and can be considered as a perturbation, our results on the dynamics of the list could be a starting point to further calculations covering for example information flow in a network.

\ack 
The model discussed here was introduced by Chaoming
Song and Dashun Wang. We are grateful to them, L\'{a}szl\'{o} Barab\'{a}si and
M\'{a}rton Karsai for discussions. This project was partially supported by
FP7 ICTeCollective Project (No. 238597) and by the Hungarian Scientific Research
Fund (OTKA, No. K100473).

\appendix
\setcounter{section}{1}
\section*{Appendix}
\subsection{Limit theorem for a geometric series of exponentially distributed random variables} \label{sec:appLTexp}
In this section we determine the probability density function of the limit random variable $X$. In the main text we defined $X_n$ and $X$ as follows
\begin{eqnarray} 
X_{n} &\sim& \sum_{k=1}^{n} e^{-k\lambda}\eta_k \\
X 	&\sim& \sum_{k=1}^{\infty} e^{-k\lambda}\eta_k  \,,
\end{eqnarray} 
where $\eta_k\sim Exp(c)$ are i.i.d. exponential random variables.
The probability density function of sums of independent exponential random variables with different parameters can be expressed \cite{Balazs_sumexp}. Substituting the proper parameters into that formula one gets
\begin{equation}
f_{X_n}(x)=\prod_{j=1}^{n}{ce^{j\lambda}}\sum_{j=1}^{n}\frac{e^{-c e^{j\lambda}x}}{\prod_{k=1,k\neq j}^{n}(e^{k\lambda}-e^{j\lambda})} \,.
\end{equation}
With some trivial algebraic manipulations this can be written in the following form
\begin{equation}
f_{X_n}(x)=c \sum_{j=1}^{n}(-1)^{j-1}\frac{e^{-j(j-3)\frac{\lambda}{2}}}{\prod_{k=1}^{j-1}(1-e^{-k\lambda})\prod_{k=1}^{n-j}(1-e^{-k\lambda})}e^{-c e^{j\lambda}x} \,.
\end{equation}
The probability density function in the $n\rightarrow \infty$ limit reads
\begin{equation} \label{eq:Xdensity}
f_{X}(x)=\frac{c}{\prod_{k=1}^{\infty}(1-e^{-k\lambda})} \sum_{j=1}^{\infty}(-1)^{j-1}\frac{e^{-j(j-3)\frac{\lambda}{2}}}{\prod_{k=1}^{j-1}(1-e^{-k\lambda})}e^{-c e^{j\lambda}x} \,.
\end{equation}
The products $\prod_{k=1}^{s}(1-e^{-k\lambda})$ are usually cited as q-Pochhammer symbols $(e^{-\lambda},e^{-\lambda})_{s}$ and are convergent in the $s \rightarrow \infty$ limit. Equation \eref{eq:Xdensity} shows clearly that the asymptotic behaviour of the limit random variable $X$ is determined by the first term in the sum.   

\subsection{Central limit theorem for the power-law decaying case} \label{sec:appCLTpower}
The Lyapunov condition for the central limit theorem reads: 
\begin{eqnarray} \label{eq:lyapunov}
&\frac{\sum_{i=1}^n\mathbb{E} \left|\xi_i-\mathbb{E}\xi_i\right|^{2+\delta}}{(\sum_{i=1}^n\mathbb{D}^2\xi_i)^{1+\delta/2}} \rightarrow 0 \ \ \textnormal{as $n\rightarrow \infty$} \,.
\end{eqnarray}
If this condition is fulfilled for any $\delta$, then the central limit theorem can be used. We test this condition at $\delta=1$ with the following moments:
\begin{eqnarray}
&\mathbb{E}(\xi_k) = \frac{1}{Q_k}=\frac{1}{c} k^{\sigma-1}\\
&\mathbb{D}^2(\xi_k) = \frac{1}{Q_k^2}=\frac{1}{c^2} k^{2\sigma-2}\\
&\mathbb{E}(|\xi_k-\mathbb{E}\xi_k|^3) = \frac{12 - 2 e}{e Q_k^3}=\frac{12 - 2 e}{e c^3} k^{3\sigma-3} \,.
\end{eqnarray}
Substituting these and approximating the sums by integrals yield
\begin{eqnarray}
&\frac{\sum\mathbb{E} |\xi_i-\mathbb{E}\xi_i|^3}{(\sum\mathbb{D}^2\xi_i)^{3/2}}  
\sim \frac{\int_0^{n}k^{3\sigma-3}dk}{\left(\int_0^{n}k^{2\sigma-2}dk\right)^{3/2}} \sim n^{-1/2}\rightarrow 0 & \ \ \textnormal{as $n\rightarrow \infty$} \,.
\end{eqnarray}

\subsection{Central limit theorem for the stretched exponential case ($\nu<1$)} \label{sec:appCLTstrexp}
We test the Lyapunov condition \eref{eq:lyapunov} at $\delta=1$ similarly to the previous case. The moments are the following:
\begin{eqnarray}
&\mathbb{E}(\xi_k) = \frac{1}{Q_k}=\frac{1}{c} e^{\lambda k^{\nu}}\\
&\mathbb{D}^2(\xi_k) = \frac{1}{Q_k^2}=\frac{1}{c^2} e^{2\lambda k^{\nu}}\\
&\mathbb{E}(|\xi_k-\mathbb{E}\xi_k|^3) = \frac{12 - 2 e}{e Q_k^3}=\frac{12 - 2 e}{e c^3} e^{3\lambda k^{\nu}} \,.
\end{eqnarray}
We make integral approximation of the sums of the moments above:
\begin{equation}\sum_{k=1}^{n}{e^{a \lambda k^{\nu}}}\approx \int_{0}^{n}{e^{a \lambda k^{\nu}}} dk = \frac{1}{\nu \lambda^{1/\nu}}\int_{0}^{a \lambda n^{\nu}}e^{y} y^{1/\nu-1} \approx \frac{1}{a\lambda \nu}e^{a \lambda n^{\nu}} n^{1-\nu} .
\end{equation}
The last approximation comes from the property that $\int_{0}^{a} e^{y} y^{1/\nu-1} \stackrel{a\rightarrow \infty}{\approx} e^{a} a^{1/\nu-1} $. 
With these we have:
\begin{eqnarray}
&\frac{\sum\mathbb{E} |\xi_i-\mathbb{E}\xi_i|^3}{(\sum\mathbb{D}^2\xi_i)^{3/2}}  
\sim \frac{e^{3\lambda n^{\nu}}n^{1-\nu}}{(e^{2\lambda n^{\nu}}n^{1-\nu})^{3/2}}
=n^{-\frac{1-\nu}{2}} \rightarrow 0 & \ \ \textnormal{as $n\rightarrow \infty$}
\end{eqnarray}
if $\nu<1$.

Now we give some more details in calculating the $\beta$ exponent. 
The mean and variance appearing in the central limit theorem are 
\begin{eqnarray}
\sum_{i=1}^{n-1} \mathbb{E} \xi_i &\approx \frac{1}{c \lambda \nu}e^{\lambda n^{\nu}}n^{1-\nu}\\ 
\sum_{i=1}^{n-1} \mathbb{D}^2 \xi_i& \approx \frac{1}{2 c^2 \lambda \nu} e^{2\lambda n^{\nu}}n^{1-\nu} \,,
\end{eqnarray}
and the approximating probability density function is
\begin{equation}
q(n,t) \approx \sqrt{\frac{c^2 \lambda \nu}{\pi}} e^{-\lambda n^{\nu}}n^{-\frac{1-\nu}{2}}
\exp\left\{-\frac{\left(t-e^{\lambda n^{\nu}}n^{1-\nu}/(c \lambda \nu)\right)^2}{2 \frac{1}{2 c^2\lambda \nu} e^{2\lambda n^{\nu}}n^{1-\nu}}\right\}  \,.
\end{equation}
We introduce a new variable 
\begin{eqnarray}
y&=& n^{\frac{1-\nu}{2}}-c \lambda \nu t e^{-\lambda n^{\nu}}n^{\frac{\nu-1}{2}} \\
dy&=&\left[\frac{1-\nu}{2}n^{\frac{-\nu+1}{2}}+(n^{\frac{1-\nu}{2}}-y)(\lambda \nu n^{\nu-1}+\frac{1-\nu}{2}n^{-1})\right]dn \,,
\end{eqnarray}
in the second equation $n\equiv n(y)= (\frac{1}{\lambda}\log(c\lambda \nu t))^{1/\nu}+\ordo(\log(c\lambda \nu t)^{1/\nu})+y \, \ordo(\log(c\lambda \nu t)^{1/\nu})$.
The dominant term in the Jacobian when $t\rightarrow \infty$ is $\lambda \nu n^{\frac{\nu-1}{2}}$.
With these the integral $\int_{0}^{\infty}q(n,t) dn$  reads 
\begin{small}
\begin{eqnarray}
\sqrt{\frac{c^2 \lambda \nu}{\pi}} \int_{-\infty}^{\infty} e^{-\frac{1}{\lambda \nu} y^2}  \left(\underline{\frac{n^{\frac{1-\nu}{2}}}{c \lambda \nu t}}-\frac{y}{c \lambda \nu t}\right) \cdot \\ \nonumber
\qquad \cdot \left[\frac{1-\nu}{2}n^{\frac{-\nu+1}{2}}+(\underline{n^{\frac{1-\nu}{2}}}-y)(\underline{\lambda \nu n^{\nu-1}}+\frac{1-\nu}{2}n^{-1})\right]^{-1} \, dy =  \\
  = \sqrt{\frac{\lambda \nu}{\pi}}\frac{1}{\lambda^{1+1/\nu} \nu^2} \int_{-\infty}^{\infty} e^{-\frac{1}{\lambda \nu} y^2}(\frac{1}{t \log(c\lambda\nu t)^{1-1/\nu}}+\ordo(\frac{1}{t \log(c\lambda\nu t)^{1-1/\nu}})) \, dy \\
  =\frac{1}{\lambda^{1/\nu}\nu} \frac{1}{t \log(c\lambda\nu t)^{1-1/\nu}}+\ordo(\frac{1}{t \log(c\lambda\nu t)^{1-1/\nu}})
\end{eqnarray}
\end{small}
We underlined the terms that determine the decay of the integral for large $t$.

\end{document}